\documentclass[preprint,aps,showpacs]{revtex4}
\usepackage{amsmath}
\usepackage{amssymb}
\usepackage{graphicx}
\usepackage{dcolumn}

\begin{document}

\title{Nonlinear transport and oscillating
magnetoresistance in double quantum wells}
\author{N. C. Mamani$^1$, G. M. Gusev$^1$, O. E. Raichev$^{1,2}$,
T. E. Lamas$^1$, and A. K. Bakarov$^{1,3}$}
\affiliation{$^1$Instituto de F\'{\i}sica da Universidade de S\~ao
Paulo, CP 66318 CEP 05315-970, S\~ao Paulo, SP, Brazil \\
$^2$Institute of Semiconductor Physics, NAS of Ukraine, Prospekt
Nauki 45, 03028, Kiev, Ukraine \\
$^3$Institute of Semiconductor Physics, Russian Academy
of Sciences, Novosibirsk 630090, Russia}

\date{\today}

\begin{abstract}

We study the evolution of low-temperature magnetoresistance in
double quantum wells in the region below 1 Tesla as the applied
current density increases. A flip of the magneto-intersubband
oscillation peaks, which occurs as a result of the current-induced
inversion of the quantum component of resistivity, is observed. We
also see splitting of these peaks as another manifestation of
nonlinear behavior, specific for the two-subband electron systems.
The experimental results are quantitatively explained by the theory
based on the kinetic equation for the isotropic non-equilibrium part
of electron distribution function. The inelastic scattering time is
determined from the dependence of the inversion magnetic field on
the current.

\pacs{73.23.-b, 73.43.Qt, 73.50.Fq}

\end{abstract}

\maketitle

\section{Introduction}

The nonlinear transport of electrons in two-dimensional (2D)
electron systems placed in a perpendicular magnetic field has been
extensively studied in the past in connection with the breakdown of
the quantum Hall effect at high current densities.$^1$ More
recently, it was realized that the current causes substantial
modifications of the resistance even in the region of weak magnetic
fields and relatively high temperatures, when the Landau levels are
thermally mixed so the Shubnikov-de Haas oscillations (SdHO) are
suppressed.

The present interest to the static (dc) nonlinear transport in 2D
systems is stimulated by observation of two important phenomena.
First, in high-mobility systems there appears a special kind of
magnetotransport oscillations, when the resistance oscillates as a
function of either magnetic field or electric current.$^{2-4}$
Second, it is found that the current substantially decreases the
resistance even at moderate applied voltages.$^{3,5}$ The observed
phenomena are of quantum origin, they are caused by the Landau
quantization of electron states and reflect the influence of the
current on the quantum contribution to resistivity. The oscillating
behavior is explained by modification of the electron spectrum in
the presence of high Hall field,$^{2,3,6}$ while the decrease of the
resistance is most possibly governed by modification of electron
diffusion in the energy space, which leads to the oscillating
non-equilibrium contribution to the distribution function of
electrons.$^{7}$ A theory describing both these phenomena in a
unified way has been recently presented.$^8$

In contrast to the Hall field-induced resistance oscillations, the
phenomenon of decreasing resistance has not been studied extensively
in experiment. Though the available data$^5$ support the
theory$^{7,8}$ predicting nontrivial changes in the distribution
function as a result of dc excitation under magnetic fields, they
are not sufficient for definite interpretation of the observed
phenomenon in terms of this theory. For better understanding of the
physical mechanisms of nonlinear behavior, further investigations
are necessary.

In this paper, we undertake the studies of nonlinear
magnetotransport in double quantum wells (DQWs), which are
representative for the systems with two closely separated occupied
2D subbands. In contrast to the quantum wells with a single occupied
subband, the positive magnetoresistance,$^9$ which originates from
the Landau quantization, is modulated in DQWs by the
magneto-intersubband (MIS) oscillations.$^{10}$ These oscillations,
whose maxima correspond to integer ratios of the subband splitting
energy $\Delta_{12}$ to the cyclotron energy $\hbar \omega_c$, are
caused by periodic variation of the probability of elastic
intersubband scattering of electrons by the magnetic field as the
density of electron states becomes an oscillating function of
energy. As a result, the changes in the quantum contribution to the
conductivity are directly seen from the corresponding changes of the
MIS oscillation amplitudes. In particular, we observe a remarkable
manifestation of nonlinearity in DQWs, the inversion of the MIS
oscillation picture, which appears when the quantum
magnetoresistance changes from positive to negative as a result of
increased current (Fig. 1). By adopting the ideas of the theory of
Ref. 7, we explain basic features of our experimental data and
determine the inelastic relaxation time of electrons in our samples.

The paper is organized as follows. In Sec. II we describe the
experimental details and present the results of our measurements. In
Sec. III we generalize the theory of Ref. 7 to the case of
two-subband occupation. A discussion, including comparison of
experimental results with the results of our calculations, is given
in Sec. IV. The last section contains the concluding remarks.

\section{Experiment}

The samples are symmetrically doped GaAs double quantum wells with
equal widths $d_{W}=14$ nm separated by Al$_{x}$Ga$_{1-x}$As
barriers with width $d_{b}$=1.4, 2, and 3.1 nm. Both layers are
shunted by ohmic contacts. Over a dozen specimens of both the Hall
bars and van der Pauw geometries from three wafers have been
studied. We have studied the dependence of the resistance of
symmetric balanced GaAs DQWs on the magnetic field $B$ at different
applied voltages and temperatures. While similar results has been
obtained in all samples with different configuration and barrier
width, we focus on measurements performed on two samples with
barrier width $d_{b}$=1.4 nm. The samples have mobilities of $9.75
\times 10^{5}$ cm$^{2}$/V s (sample A) and $4.0 \times 10^{5}$
cm$^{2}$/V s (sample B) and total electron density $n_s =1.01 \times
10^{12}$ cm$^{-2}$. The samples are Hall bars of width 200 $\mu$m
and length 500 $\mu$m between the voltage probes. The resistance
$R=R_{xx}$ was measured by using the standard low-frequency lock-in
technique for low value of the current. We also use DC current,
especially for high-current measurements. The results obtained with
AC and DC techniques are similar. The subband separation
$\Delta_{12}$, found from the MIS oscillation periodicity at low
$B$, is 3.7 meV for sample A and 5.1 meV for sample B.

\begin{figure}[ht]
\begin{center}\leavevmode
\includegraphics[width=8cm]{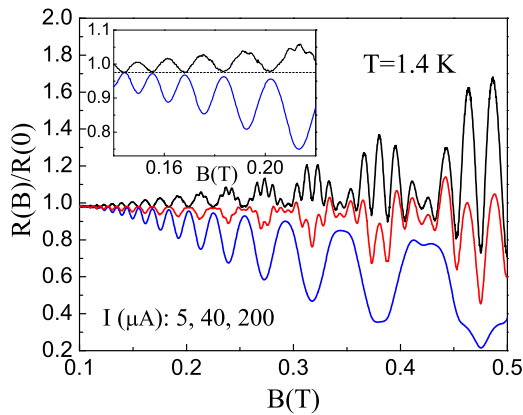}
\end{center}
\addvspace{-0.8 cm} \caption{(Color online) Magnetoresistance of the
sample A for three different currents $I$ at $T=1.4$ K. The
oscillations are inverted with the increase of the current. The
inset shows the linear and non-linear (at $I=200$ $\mu$A)
magnetoresistance in the low-field region.}
\end{figure}

The resistance of the sample A as a function of magnetic field at
different currents is presented in Figs. 1 and 2. At small currents,
the magnetoresistance is positive and modulated by the large-period
MIS oscillations clearly visible above $B=0.1$ T. The small-period
SdHO, superimposed on the MIS oscillation pattern, appear at higher
fields in the low-temperature measurements (Fig. 1). With increasing
current $I$, the amplitudes of the MIS oscillations decrease, until
a flip of the MIS oscillation picture occurs. This flip, which we
associate with inversion of the quantum component of the
magnetoresistance from positive to negative, starts from the region
of lower fields and extends to higher fields as the current
increases. Therefore, one can introduce a characteristic,
current-dependent inversion field $B_{inv}$. The inset to Fig. 2
shows the behavior of the magnetoresistance near the point of
inversion. In this point, apart from the transition from negative to
positive quantum magnetoresistance, we observe an additional feature
that looks like splitting of the MIS oscillation peaks or appearance
of the next harmonic of the MIS oscillations. This feature persists
in higher magnetic fields. In contrast to the MIS oscillations, the
SdHO are not inverted by the current, as seen in Fig. 1. However,
the SdHO amplitudes decrease as the current increases until the SdHO
completely disappear in the low-field region. We attribute this
suppression of the SdHO to electron heating at high current
densities.

\begin{figure}[ht]
\begin{center}\leavevmode
\includegraphics[width=8cm]{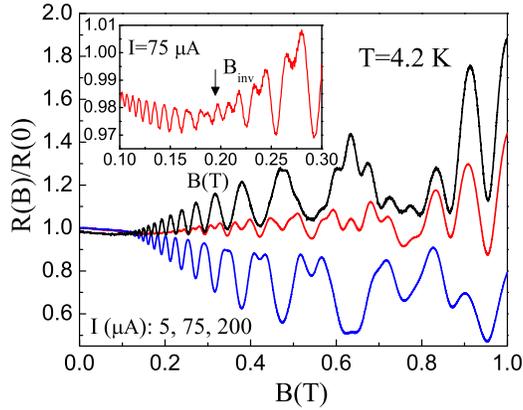}
\end{center}
\addvspace{-0.8 cm} \caption{(Color online) Magnetoresistance of the
sample A for different currents at $T=4.2$ K. The inset shows
inversion of the quantum magnetoresistance around $B=0.2$ T.}
\end{figure}

The amplitudes of inverted MIS oscillations increase with
increasing current and become larger than the MIS oscillation
amplitudes in the linear regime. At low temperatures the ratio
of the corresponding amplitudes varies between 2 and 3; see Fig. 1.
However, when the current increases further, the amplitudes of
inverted peaks slowly decrease, this decrease goes faster in the
region of lower magnetic fields. This property is seen in Figs. 3
and 4, where the magnetoresistance data for the sample B is
presented. The typical current dependence of the inverted peak
amplitudes at $T=1.4$ K is shown in the inset to Fig. 3. The
behavior of magnetoresistance at 4.2 K, shown in Fig. 4, is similar.
In the chosen interval of magnetic fields, the SdHO at 4.2 K are
suppressed even in the linear regime. The splitting of the MIS
oscillation peaks is clearly visible in Fig. 4 at $I=80$ $\mu$A. For
$I = 100$ $\mu$A this splitting apparently develops in the frequency
doubling of the MIS oscillations. Further increase of the current
suppresses this feature, leading to a more simple picture of
inverted MIS oscillations.

\begin{figure}[ht]
\begin{center}\leavevmode
\includegraphics[width=8cm]{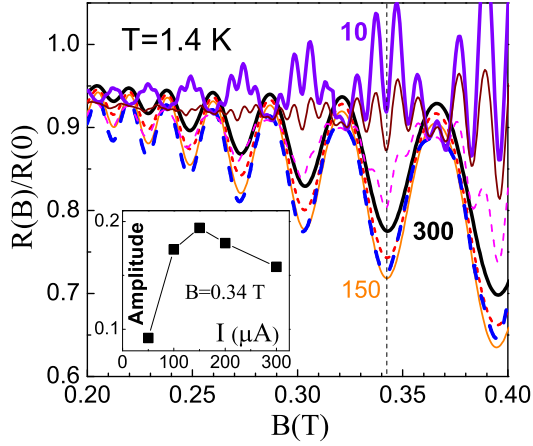}
\end{center}
\addvspace{-0.8 cm} \caption{(Color online) Magnetoresistance of the
sample B at $T=1.4$ K.  The values of the current are 10 (bold), 30,
50 (dash), 100 (bold dash), 150, 200 (short dash), and 300 (bold)
$\mu$A. The inset shows amplitudes of the inverted peaks at $B=0.34$
T.}
\end{figure}

\begin{figure}[ht]
\begin{center}\leavevmode
\includegraphics[width=8cm]{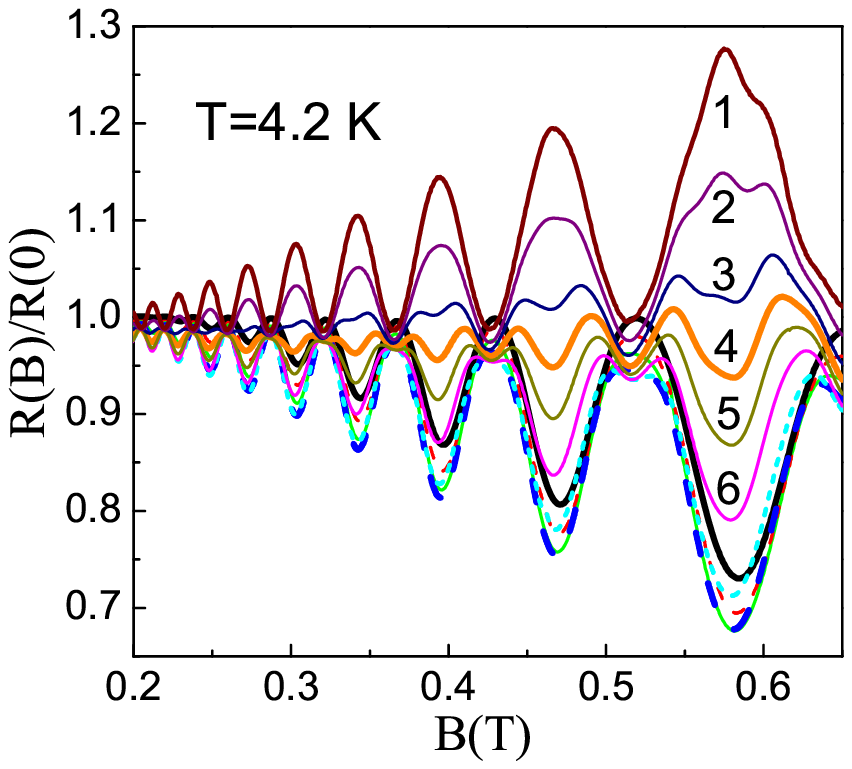}
\end{center}
\addvspace{-0.8 cm} \caption{(Color online) Magnetoresistance of the
sample B at $T=4.2$ K. The values of the current ($\mu$A) are 1, 50,
80, 100, 120, and 150 for the curves marked by the numbers from 1 to
6, respectively. The other curves corresponds to the currents of 200
(short dash), 250 (bold dash) 300 (solid), 350 (dash), and 400
(bold) $\mu$A.}
\end{figure}

\section{Theory}

The theoretical interpretation of our data is based on the physical
model of Dmitriev {\em et al.},$^{7}$ generalized to the two-subband
case. The elastic scattering of electrons is assumed to be much
stronger than the inelastic one. This scattering maintains nearly
isotropic carrier distribution at moderate currents, when the
momentum gained by an electron moving in the electric field between
the scattering events is much smaller than the Fermi momentum. Since
the intersubband elastic scattering is also much stronger than the
inelastic scattering, the isotropic part of electron distribution
function, $f_{\varepsilon}$, is common for both subbands and depends
only on the electron energy $\varepsilon$. When the current of
density $j$ flows through the sample, the kinetic equation for this
function is written as
\begin{equation}
\frac{P}{D_{\varepsilon} \sigma_d} \frac{\partial}{\partial
\varepsilon} \sigma_d (\varepsilon) \frac{\partial}{\partial
\varepsilon} f_{\varepsilon} =- J_{\varepsilon}(f),
%1
\end{equation}
where $P=j^2 \rho_d$ is the power of Joule heating (the energy
absorbed per unit time over a unit square of electron system)
expressed through the diagonal resistivity $\rho_d$, and
$D_{\varepsilon}$ is the density of states. The function $\sigma_d
(\varepsilon)$ can be written through the electron Green's
functions, which are determined by the interaction of electrons with
static disorder potential in the presence of magnetic field. The
free-electron states in the magnetic field described by the vector
potential $(0,Bx,0)$ are characterized by the quantum numbers $j$,
$n$, and $p_y$, where $j=1,2$ numbers the electron subband of the
quantum well, $n$ is the Landau level number, and $p_y$ is the
continuous momentum. Using the free-electron basis, one obtains
\begin{eqnarray}
\sigma_{d}(\varepsilon)= \frac{e^2}{2 \pi m} {\rm Re} \left[
Q_{\varepsilon}^{AR} - Q_{\varepsilon}^{AA} \right], \\
%2,3
Q_{\varepsilon}^{ss'} = \frac{2 \omega_c}{L^2}
\sum_{nn'} \sum_{jj'} \sqrt{(n+1)(n'+1)} \sum_{p_y p_y'}  \nonumber \\
\times \left< \left< G^{jj',s}_{\varepsilon}(n+1 p_y, n'+1 p'_y)
G^{j'j,s'}_{\varepsilon}(n' p'_y, n p_y) \right> \right>,
\end{eqnarray}
where $e$ is the electron charge, $m$ is the effective mass of
electron, $G^{jj',s}_{\varepsilon}$ are the retarded ($s=R$) and
advanced ($s=A$) Green's functions, and $L^2$ is the normalization
square. The Zeeman splitting is neglected, so the electrons are
assumed to be spin-degenerate. The double angular brackets in Eq.
(3) denote averaging over the random potential. In terms of the
Green's functions, the density of states is given by
\begin{equation}
D_{\varepsilon}=\frac{2}{\pi L^2} \sum_{j n p_y} {\rm Im} \left<
\left< G^{jj,A}_{\varepsilon}(n p_y,n p_y) \right> \right>= \frac{2
m}{\pi \hbar^2} \sum_j {\rm Im} S_{j\varepsilon}.
%4
\end{equation}
The dimensionless function $S_{j\varepsilon}$ is found from the
implicit equation
\begin{eqnarray}
S_{j\varepsilon}=\frac{\hbar \omega_c}{2 \pi} \sum_{n}
\frac{1}{\varepsilon-\hbar \omega_c(n+1/2)-\varepsilon_j-\Sigma_{j
\varepsilon} }, \\
\Sigma_{j \varepsilon}=\sum_{j'} \frac{\hbar}{\tau_{jj'}}
S_{j'\varepsilon},~~~~~~~~~ \nonumber
%5
\end{eqnarray}
where $\omega_c$ is the cyclotron energy, $\varepsilon_j$ is the
subband energy, and $\tau_{jj'}$ are the quantum lifetimes of
electrons with respect to intrasubband ($j'=j$) and intersubband
($j' \neq j$) scattering. Equation (5) is valid when the correlation
length of the disorder potential is smaller than the magnetic
length, and the disorder-induced energy broadening of the subbands
is smaller than the subband separation $\Delta_{12}=\varepsilon_2-
\varepsilon_1$. It corresponds to the the self-consistent Born
approximation (SCBA).

According to the definition (2), the diagonal conductivity is
\begin{equation}
\sigma_d=\int d \varepsilon \left(-\frac{\partial
f_{\varepsilon}}{\partial \varepsilon} \right)
\sigma_{d}(\varepsilon).
%6
\end{equation}
Therefore, multiplying the kinetic equation (1) by the density of
states $D_{\varepsilon}$ and energy $\varepsilon$, and integrating
it over $\varepsilon$, one obtains the balance equation $P=P_{ph}$,
where $P_{ph}=-\int d \varepsilon ~\! \varepsilon D_{\varepsilon}
J_{\varepsilon}(f)$ is the power lost to the lattice vibrations
(phonons).

Below we consider the case of classically strong magnetic field,
$\omega_c \tau_{tr} \gg 1$, when $\sigma_{d}(\varepsilon)$ is
written in terms of $S_{j\varepsilon}$ as
\begin{equation}
\sigma_{d}(\varepsilon)=\frac{4e^2}{m \omega_c^2}
\left[\frac{n_1}{\tau^{tr}_{11}} ({\rm Im} S_{1 \varepsilon})^2 +
\frac{n_2}{\tau^{tr}_{22}} ({\rm Im} S_{2 \varepsilon})^2
+\frac{n_s}{\tau^{tr}_{12}} {\rm Im} S_{1 \varepsilon} {\rm Im} S_{2
\varepsilon} \right],
%7
\end{equation}
where $n_1$ and $n_2$ are the electron densities in the subbands,
$n_s=n_1+n_2$, and $\tau^{tr}_{jj'}$ are the transport times of
electrons. Both $\tau_{jj'}$ and $\tau^{tr}_{jj'}$ are determined by
the expressions
\begin{equation}
\begin{array}{c} 1/\tau_{jj'} \\
1/\tau^{tr}_{jj'} \end{array}
 \left\} = \frac{m}{\hbar^3} \int_0^{2 \pi}
\frac{d \theta}{2 \pi} w_{jj'} \left( \sqrt{ (k^2_{j} +
k^2_{j'})F_{jj'}(\theta)} \right) \times \right\{\begin{array}{c} 1 \\
F_{jj'}(\theta) \end{array} ,
%8
\end{equation}
where $w_{jj'}(q)$ are the Fourier transforms of the correlators of
the scattering potential, $F_{jj'}(\theta)=1 - 2 k_j k_{j'} \cos
\theta/(k_j^2 + k_{j'}^2)$, and $k_j$ is the Fermi wavenumber for
the subband $j$. The electron densities in the subbands are
expressed as $n_{j}=k^2_{j}/2 \pi$.

In DQWs, where the energy separation between the subbands is usually
small compared to the Fermi energy, the difference $k^2_{1}-k^2_{2}$
is small in comparison with $k^2_{1}+k^2_{2}$ so that $n_1 \simeq
n_2 \simeq n_s/2$. Furthermore, in the symmetric (balanced) DQWs,
where the electron wave functions are delocalized over the layers
and represent themselves symmetric and antisymmetric combinations of
single-layer orbitals, one has nearly equal probabilities for
intrasubband and intersubband scattering owing to $w_{11}(q) \simeq
w_{22}(q) \simeq w_{12}(q)$, provided that interlayer correlation of
the scattering potentials is weak. Therefore, $\tau_{jj} \simeq
\tau_{12} \simeq 2 \tau$, and $\tau^{tr}_{jj} \simeq \tau^{tr}_{12}
\simeq 2 \tau_{tr}$, where $\tau$ and $\tau_{tr}$ are the averaged
quantum lifetime and transport time, respectively.
%$1/\tau=(1/\tau_{11}+1/\tau_{22})/2+1/\tau_{12}$
%$1/\tau_{tr}=(1/\tau^{tr}_{11}+1/\tau^{tr}_{22})/2+1/\tau^{tr}_{12}$
In these approximations, Eq. (7) is written in the most simple way:
\begin{equation}
\sigma_d (\varepsilon) \simeq \sigma^{(0)}_d {\cal
D}^2_{\varepsilon},~~~~ {\cal D}_{\varepsilon}=\frac{1}{2}({\cal
D}_{1\varepsilon}+{\cal D}_{2\varepsilon}), ~~~~{\cal D}_{j
\varepsilon} = 2 {\rm Im} S_{j \varepsilon}
%9
\end{equation}
where $\sigma^{(0)}_d=\sigma^2_{\bot} \rho_0$,
$\sigma_{\bot}=e^2n_s/ m \omega_c$ is the Hall conductivity, and
$\rho_0=m/e^2 \tau_{tr} n_s$ is the classical resistivity. The
function ${\cal D}_{\varepsilon}= 1 + \gamma_{\varepsilon}$ is the
dimensionless density of states, containing oscillating (periodic in
$\hbar \omega_c$) part $\gamma_{\varepsilon}$. Therefore, it is
convenient to solve the kinetic equation by representing the
distribution function as a sum $f^{0}_{\varepsilon}+ \delta
f_{\varepsilon}$, where the first term slowly varies on the scale of
cyclotron energy, while the second one rapidly oscillates.$^{7}$ The
first term satisfies the equation
\begin{equation}
\kappa \frac{\partial^2}{\partial \varepsilon^2} f^{0}_{\varepsilon}
=- J_{\varepsilon}(f^{0}),~~~\kappa=\frac{\pi \hbar^2 j^2
\rho_0}{2m}.
%10
\end{equation}
Solution of this equation can be satisfactory approximated by a
heated Fermi distribution. This is always true if the
electron-electron scattering dominates over the electron-phonon
scattering and over the electric-field effect described by the
left-hand side of Eq. (10). In this case, the Fermi distribution of
electrons is maintained against the field-induced diffusion in the
energy space, while the electron-phonon scattering determines the
effective electron temperature $T_e$. In the general case, a
numerical solution of Eq. (10) involving electron-phonon scattering
in the collision integral$^{11}$ confirms that $f^{0}_{\varepsilon}$
is very close to the heated Fermi distribution.

The equation for the oscillating part, $\delta f_{\varepsilon}$, is
then written in the following form:
\begin{equation}
{\cal D}_{\varepsilon} \frac{\partial^2}{\partial \varepsilon^2}
\delta f_{\varepsilon} + 2 \frac{\partial {\cal
D}_{\varepsilon}}{\partial \varepsilon} \frac{\partial}{\partial
\varepsilon} \delta f_{\varepsilon} + \kappa^{-1}
J_{\varepsilon}(\delta f) = -2 \frac{\partial {\cal
D}_{\varepsilon}}{\partial \varepsilon} \frac{\partial
f^{0}_{\varepsilon}}{\partial \varepsilon} .
%11
\end{equation}
Below we search for the function $\delta f_{\varepsilon}$ in the
form $\delta f_{\varepsilon}=(\partial f^{0}_{\varepsilon}/\partial
\varepsilon) \varphi_{\varepsilon}$, where $\varphi_{\varepsilon}$
is a periodic function of energy. Taking into account that the main
mechanism of relaxation of the distribution $\delta f_{\varepsilon}$
is the electron-electron scattering, one may represent the
linearized collision integral $J_{\varepsilon}(\delta f)$ as
\begin{eqnarray}
J_{\varepsilon}(\delta f) = - \frac{1}{\tau_{in}} \frac{\partial
f^{0}_{\varepsilon}}{\partial \varepsilon}  \frac{1}{ {\cal N} {\cal
D}_{ \varepsilon}} \sum_{j j' j_1 j'_1} M_{jj',j_1 j'_1} \left<
{\cal D}_{j \varepsilon} {\cal D}_{j_1 \varepsilon+ \delta
\varepsilon} {\cal D}_{j' \varepsilon'} {\cal D}_{j'_1 \varepsilon'
- \delta \varepsilon} \right. \nonumber \\
\left. \times [\varphi_{\varepsilon} + \varphi_{\varepsilon'}-
\varphi_{\varepsilon + \delta \varepsilon}  -\varphi_{\varepsilon' -
\delta \varepsilon} ] \right>_{\varepsilon', \delta
\varepsilon},~~~~{\cal N}=\sum_{j j' j_1 j'_1} M_{jj',j_1 j'_1},
%12
\end{eqnarray}
where $\delta \varepsilon$ is the energy transferred in
electron-electron collisions, $M_{jj',j_1 j'_1}$ is the probability
of scattering (when electrons from the states $j$ and $j'$ come to
the states $j_1$ and $j'_1$), ${\cal N}$ is the normalization
constant, and the angular brackets $\left< \ldots
\right>_{\varepsilon', \delta \varepsilon}$ denote averaging over
the energies $\varepsilon'$ and $\delta \varepsilon$. Expression
(12) is a straightforward generalization of the result of Ref. 7.
The characteristic inelastic scattering time $\tau_{in}$ describes
the relaxation at low magnetic fields, when ${\cal D}_{j
\varepsilon}$ are close to unity. In this case the collision
integral acquires the most simple form $J_{\varepsilon}(\delta f) =
-\delta f_{\varepsilon}/\tau_{in}$, i.e. the relaxation time
approximation is justified.

The resistivity $\rho_d = \sigma^{(0)}_d/\sigma^2_{\bot}$ is
written, according to Eq. (6), in the form
\begin{equation}
\rho_{d}= \rho_0 \int d \varepsilon {\cal D}^2_{\varepsilon} \left(
-\frac{\partial f^{0}_{\varepsilon}}{\partial \varepsilon} \right)
\left(1+ \frac{\partial \varphi_{\varepsilon}}{\partial \varepsilon}
\right),
%13
\end{equation}
where we have taken into account that $\partial
f_{\varepsilon}/\partial \varepsilon \simeq (\partial
f^{0}_{\varepsilon}/\partial \varepsilon ) \left[1+ \partial
\varphi_{\varepsilon}/\partial \varepsilon \right]$. Therefore, in
order to calculate the resistivity, one should find
$\varphi_{\varepsilon}$ by using Eqs. (11) and (12). In general, Eq.
(12) is an integro-differential equation that cannot be solved
analytically. However, the property of periodicity allows one to
expand $\varphi_{\varepsilon}$ in series of harmonics,
$\varphi_{\varepsilon} =\sum_{k} \varphi_k \exp(2 \pi i k
\varepsilon/\hbar \omega_c)$, and represent Eq. (11) as a system of
linear equations:
\begin{equation}
(Q^{-1}+k^2) \varphi_k + \sum_{k'=-\infty}^{\infty}\left[
(2kk'-k'^2) \gamma_{k-k'}+Q^{-1} C_{kk'} \right] \varphi_{k'} = 2 i
k \frac{\hbar \omega_c}{2 \pi} \gamma_k,
%14
\end{equation}
where
\begin{equation}
Q= \frac{2 \pi^3 j^2}{e^2 n_s \omega^2_c}\frac{\tau_{in}}{\tau_{tr}}
%15
\end{equation}
is a dimensionless parameter characterizing the nonlinear effect of
the current on the transport. The matrix $C_{kk'}$, whose explicit
form is not shown here, describes the effects of electron-electron
scattering beyond the relaxation time approximation.

The harmonics of the density of states, $\gamma_k$, as well as the
coefficients $C_{kk'}$, which are expressed in terms of products of
these harmonics, are proportional to the Dingle factors $\exp(-k
\pi/\omega_c \tau)$. Therefore, searching for the coefficients
$\varphi_k$ at weak enough magnetic fields, when $e^{-\pi/\omega_c
\tau}$ is small, one can take into account only a single ($k=\pm 1$)
harmonic. Within this accuracy, one should also neglect the sum in
Eq. (14). This leads to a simple solution $\varphi_{\pm 1}=\pm i
\gamma_{\pm 1}(\hbar \omega_c/\pi) Q/(1+Q)$. Since $\gamma_{+1} +
\gamma_{-1}=-2 e^{-\pi/\omega_c \tau} \cos(\pi \Delta_{12}/\hbar
\omega_c)$, Eq. (13) is reduced to a simple analytical expression
for the resistivity:
\begin{eqnarray}
\frac{\rho_{d}}{\rho_0}= 1 + e^{-2 \pi/\omega_c \tau}
\frac{1-3Q}{1+Q}\left(1+\cos \frac{2 \pi \Delta_{12}}{\hbar
\omega_c}\right) \nonumber \\
-4 e^{-\pi/\omega_c \tau} {\cal T} \cos \left( \frac{2 \pi
\varepsilon_F}{\hbar \omega_c} \right) \cos \left( \frac{\pi
\Delta_{12}}{\hbar \omega_c} \right).
%16
\end{eqnarray}
The second term in this expression, proportional to $e^{-2
\pi/\omega_c \tau}$, differs from a similar term of the
single-subband theory$^7$ by the modulation factor $\left[1+\cos (2
\pi \Delta_{12}/\hbar \omega_c) \right]/2$ describing the MIS
oscillations. The last term in Eq. (16) describes the SdHO, which
are thermally suppressed because of the factor ${\cal T}=(2 \pi^2
T_e/\hbar \omega_c)/\sinh(2 \pi^2 T_e/\hbar \omega_c)$. The Fermi
energy $\varepsilon_F$ is counted from the middle point between the
subbands, $(\varepsilon_1+\varepsilon_2)/2$, and, therefore, is
directly proportional to the total electron density, $\varepsilon_F=
\hbar^2 \pi n_s/2m$.

\section{Results and discussion}

The basic features of our experimental findings can be understood
within Eqs. (16) and (15). In the linear regime, when the parameter
$Q$ is small, this equation gives a good description of the MIS
oscillations experimentally investigated in Ref. 10. As the current
increases, the amplitudes of these oscillations decrease, and then
the flip occurs, when the MIS peaks become inverted. In contrast,
the SdHO peaks are not affected by the the current directly, and
their decrease is caused by the effect of heating. The flip of the
MIS oscillations corresponds to $Q=1/3$. Since $Q$ is inversely
proportional to the square of the magnetic field, there exists the
inversion field, $B_{inv}$, determined from the equation $Q=1/3$,
where $Q$ is given by Eq. (15). This feature is observed in our
experiment, see the inset to Fig. 2. For the sample B, we have
extracted $B_{inv}$ for several values of the current. The results
are shown in Fig. 5. At 4.2 K the experimental points follow the
linear $B_{inv}(I)$ dependence predicted by Eq. (15). Since the
ratio $B_{inv}/I$ is proportional to the square root of the
inelastic relaxation time $\tau_{in}$, we are able to estimate this
time from experimental data as $\tau_{in} \simeq 64$ ps at $T=4.2$
K. Assuming the $T^{-2}$ scaling of this time,$^{7}$ one obtains
$\hbar/\tau_{in}=6.6$ mK at $T=1$ K, which is not far than
the theoretical estimate $\hbar/\tau_{in} =4$ mK at $T=1$ K based
on the consideration of electron-electron scattering.$^{7}$
The positions of experimental points at $T=1.4$ K also fit within
this picture if the electron heating is taken into account. The
increase of electron temperature with increasing current (heating
effect) leads to deviation of the $B_{inv}(I)$ dependence from
linearity because of temperature dependence of $\tau_{in}$, and this
deviation is essential at $T=1.4$ K; see Fig. 5. The same consideration,
applied to the high-mobility sample A, gives the inelastic scattering
time $\tau_{in} \simeq 108$ ps at $T=4.2$ K, which is very close to
the theoretical estimate.

\begin{figure}[ht]
\begin{center}\leavevmode
\includegraphics[width=8cm]{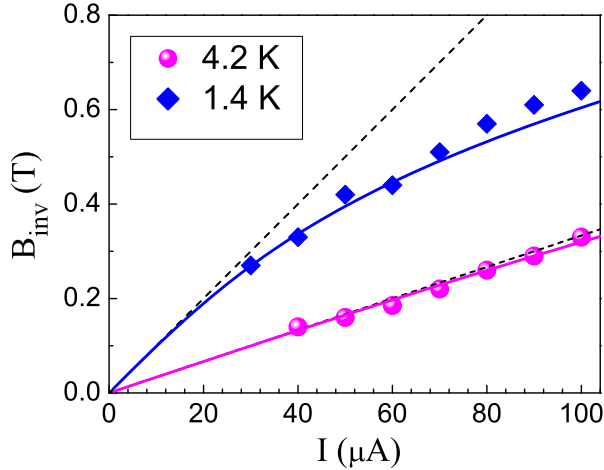}
\end{center}
\addvspace{-0.8 cm} \caption{(Color online) Dependence of the
inversion field on the current for the sample B at $T=4.2$ K and
$T=1.4$ K. (points) The dashed lines correspond to a linear
$B_{inv}(I)$ dependence assuming $\tau_{in}= 64$ ps at 4.2 K (580
ps at 1.4 K). The solid lines represent the calculated $B_{inv}(I)$
dependence taking into account electron heating by the current.}
\end{figure}

When the current becomes high enough ($Q \gg 1$), Eq. (16) predicts
saturation of the resistance, when the amplitudes of inverted MIS
peaks are three times larger than the amplitudes of the MIS peaks in
the linear regime ($Q \ll 1$). We indeed observe the regime resembling
a saturation, with almost three times increase in the amplitudes of
inverted peaks for both samples at $T=1.4$ K (see Figs. 1 and 3).
For higher temperatures the behavior is similar, though the maximum
amplitudes of inverted peaks are only slightly larger than the amplitudes
in the linear regime. We explain this by the effect of heating on the
characteristic times. Though the resistivity in the high-current regime
($Q \gg 1$) no longer depends on $\tau_{in}$, there is a sizeable
decrease in the quantum lifetime $\tau$ with increasing temperature,$^{10}$
which takes place because the electron-electron scattering contributes
into $\tau$. As a result, the Dingle factor decreases, and the quantum
contribution to the resistance becomes smaller as the electrons are heated.
At higher initial temperature, when $\tau_{in}$ is smaller, the regime
$Q \gg 1$ requires higher currents. The corresponding increase in
heating reduces the quantum contribution, so the maximum amplitudes of
inverted peaks never reach the theoretical limit and are expected to
decrease with increasing initial temperature. The slow suppression
of the inverted peaks with further increase in the current (see the inset to
Fig. 3) is explained by the same mechanism. This conclusion is supported by
the experimental observation that the suppression is more efficient at lower
magnetic fields, when the Dingle factor $\exp(-\pi/\omega_c \tau)$ is more
sensitive to the temperature dependence of quantum lifetime $\tau$.

\begin{figure}[ht]
\begin{center}\leavevmode
\includegraphics[width=8cm]{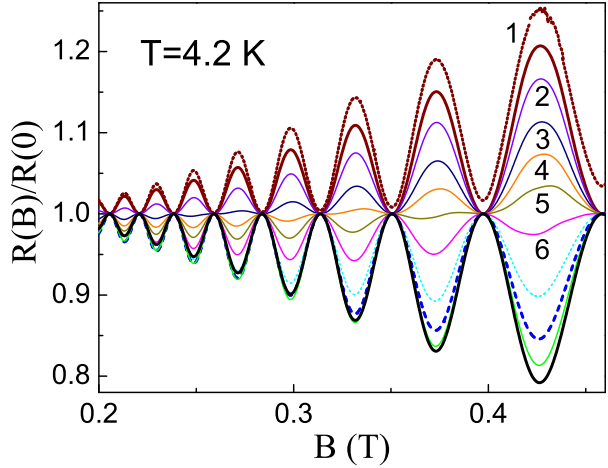}
\end{center}
\addvspace{-0.8 cm} \caption{(Color online) Calculated
magnetoresistance of the sample B at $T=4.2$ K and different
currents: 1, 50, 80, 100, 120, and 150 $\mu$A for the curves marked
by the numbers from 1 to 6; the other curves corresponds to $I=200$
(short dash), 250 (bold dash), 300 (solid), and 400 (bold) $\mu$A.
The additional (dashed) line 1 shows the linear magnetoresistance
determined by the SCBA calculation of the density of states in Eq.
(13).}
\end{figure}

To illustrate the above-discussed relation of the basic theoretical
predictions to our experiment, we present the results of theoretical
calculations according to Eqs. (15) and (16) in Fig. 6. The
calculations are done for the sample B at 4.2 K, so the theoretical
curves show the expected behavior of the measured magnetoresistance
from Fig. 4. We take into account the effect of heating, described
by using the collision integral for interaction of electrons with
acoustic phonons$^{11}$ and temperature dependence of the quantum
lifetime $\tau$ of electrons determined empirically from the studies
of the MIS oscillations in the linear regime.$^{10}$ The theoretical
plots demonstrate a reasonable qualitative agreement with the
experiment. However, the theory predicts a slower suppression of the
inverted peak amplitudes with increasing current at weak magnetic fields. 
This may be a consequence of underestimated heating,$^{12}$ because the 
screening effect on the electron-phonon interaction$^{13}$ has not been 
taken into account in the calculation of the power loss to acoustic
phonons. Similar calculations carried out for different samples at
different temperatures are also in agreement with experimental data.

\begin{figure}[ht]
\begin{center}\leavevmode
\includegraphics[width=8cm]{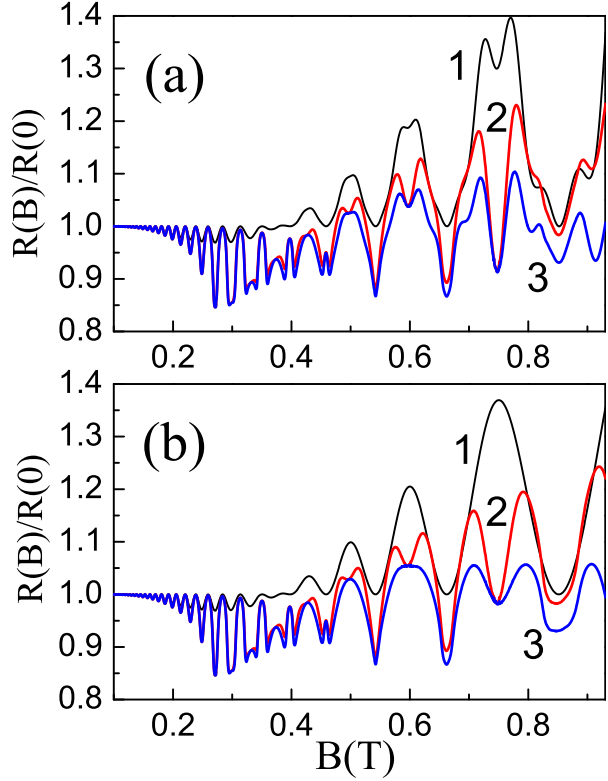}
\end{center}
\addvspace{-0.8 cm} \caption{(Color online) (a) Calculated
magnetoresistance of the sample B at $T=4.2$ K and $I=120$ $\mu$A.
The plot 1 correspond to simple theory [Eq. (16)], while the others
represent the results of numerical solution of Eq. (14) for the
cases of subband-independent electron-electron scattering (2) and
only intrasubband electron-electron scattering (3). (b) The same
plots, where the SdHO contribution is excluded. The density of
states is found within the SCBA.}
\end{figure}

The simple theory fails do describe the interesting and unexpected
feature observed in our experiment, the current-induced splitting of
the MIS oscillation peaks. This kind of nonlinear behavior is
well-reproducible, we see it in different samples. We have found
that a possible explanation of this feature can be based on the
theory presented in Sec. III, if higher harmonics of the
distribution function $\delta f_{\varepsilon}$ are taken into
account. We have carried out a numerical solution of the system of
equations (14) under some simplifying assumptions about the
collision integral. In the first case, we have assumed equal
probabilities for all possible electron-electron scattering
processes, so the matrix $M_{jj',j_1 j'_1}$ in Eq. (12) is replaced
by a constant. Another limiting case we consider is the complete
neglect of intersubband transitions in electron-electron collisions,
when $M_{jj',j_1 j'_1} \propto \delta_{jj_1} \delta_{j'j'_1}$. This
case is also reasonable, since electron-electron scattering at low
temperatures assumes a small momentum transfer, so the intersubband
scattering contribution should be suppressed owing to reduction of
the overlap integrals of envelope wave functions of electrons. Then,
the coefficients $\gamma_k$ and $C_{kk'}$ have been determined by
using the density of states numerically calculated within the SCBA;
see Eq. (5). The results, corresponding to $I=120$ $\mu$A for the
sample B are presented in Fig. 7. In the low-field region, where the
MIS peaks are inverted, the calculation shows a considerable
increase in their amplitudes above 0.2 T, where contribution of
higher harmonics of the density of states becomes essential. This
enhancement occurs because of the current-induced mixing between
different harmonics of the distribution function, formally coming
from the term with $\gamma_{k-k'}$ in the sum in Eq. (14). In
contrast, in the linear regime, the SCBA magnetoresistance is close
to the magnetoresistance calculated within the single-harmonic
approximation [Eq. (16)]; see Fig. 6. Above 0.27 T, where the Landau
levels become separated, one can see features associated with the
specific semi-elliptic shape of the SCBA density of states. In the
vicinity of the inversion field ($B_{inv} \simeq 0.4$ T), where the
contribution of the first harmonic of the distribution function is
suppressed ($Q \simeq 1/3$) while the higher harmonics are still
active, two sets of MIS peaks are seen. It is not surprising,
because higher harmonics of the density of states contain the
factors $\cos(k \pi \Delta_{12}/\hbar \omega_c)$ describing higher
harmonics of the MIS oscillations. Above the inversion field, the
resistance is considerably smaller than the resistance predicted by
the single-harmonic approximation, and a splitting of the MIS peaks
occurs. The splitting increases with the increase of the magnetic
field. These effects are caused by the contribution of higher
harmonics of the density of states in the collision integral.
Indeed, in the single-harmonic approximation the collision integral
contains only the outcoming term proportional to
$\varphi_{\varepsilon}$. This approximation becomes insufficient in
higher magnetic fields, when incoming terms in the collision
integral (12) are also important, so the relaxation of the
distribution function, which counteracts the diffusion of electrons
in the energy space, becomes less efficient. This means that the
effect of the current on the distribution function increases, and
the resistance is lowered. The described suppression of the
collision-integral term is more significant in the regions of the
MIS resonances, when $\Delta_{12}/\hbar \omega$ is integer, because
the peaks of the density of states are the narrowest in these
conditions, and the energies transferred in the electron-electron
collisions, $\delta \varepsilon$, are small. Away from the MIS
resonances, the energy space for electron-electron scattering
increases, especially when the intersubband transitions are allowed
(see curve 2 in Fig. 7). Therefore, the relaxation is less
suppressed as compared to the center of the MIS peak, and the effect
of the current is weaker. The above consideration explains why the
centers of the MIS peaks drop down, so the peak splitting takes
place.

\begin{figure}[ht]
\begin{center}\leavevmode
\includegraphics[width=8cm]{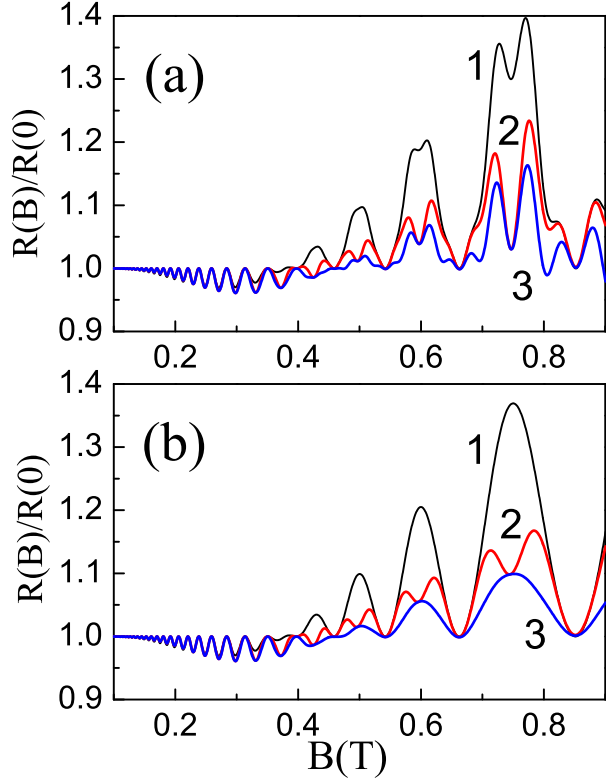}
\end{center}
\addvspace{-0.8 cm} \caption{(Color online) The same as in Fig. 7
for the Gaussian model of the density of states.}
\end{figure}

The SCBA has a limited applicability for description of the density
of electron states in the magnetic field. In particular, it leads to
non-physically sharp edges of the density of states, which generate
the harmonics $\gamma_k$ with large $k$ in Eq. (14). This apparently
leads to an overestimate of the effect of the current on the
resistance in the region where the MIS peaks are inverted, see Fig.
7. To avoid such singularities, and to have a further insight into
the problem of nonlinear magnetoresistance, we have considered the
expression
\begin{equation}
{\cal D}^{(G)}_{1,2 \varepsilon}=\frac{\hbar \omega_c}{ \sqrt{\pi}
\Gamma(\omega_c)} \sum_{n=-\infty}^{\infty} \exp \frac{[\varepsilon
\pm \Delta_{12}/2 -\hbar \omega_c(n+1/2)]^2}{\Gamma^2(\omega_c)}.
%17
\end{equation}
which corresponds to the Gaussian model for the density of states
and describes two independent sets of Landau-level peaks from each
subband (strictly speaking, the Landau-level peaks are not
independent because of elastic intersubband scattering, as follows
from Eq. (5), see more details in Ref. 14). The magnetic-field
dependence of the broadening energy $\Gamma$ has been set to make
the first [proportional to $\cos(2 \pi \varepsilon/\hbar\omega_c)$]
harmonics of ${\cal D}^{(G)}_{j \varepsilon}$ and ${\cal D}_{j
\varepsilon}$ equal. The results of the calculations using ${\cal
D}^{(G)}_{j \varepsilon}$ instead of the SCBA density of states are
shown in Fig. 8. The magnetoresistance in the region of inversion
appears to be nearly the same as predicted by the simple
single-harmonic theory. In the region above the inversion field, the
splitting of the MIS peaks does not take place if the intersubband
electron-electron scattering is forbidden. This is understandable
from the discussion given above: if different subbands contribute
into the density of states independently, the efficiency of
electron-electron collisions does not depend on the ratio
$\Delta_{12}/\hbar \omega$ and the reduction of the collision
integral owing to incoming terms causes just a uniform suppression
of the whole MIS peak. In the SCBA, when the shape of ${\cal D}_{j
\varepsilon}$ depends on this ratio, the splitting of the MIS peaks
does not necessarily require the intersubband electron-electron
scattering.

\begin{figure}[ht]
\begin{center}\leavevmode
\includegraphics[width=8cm]{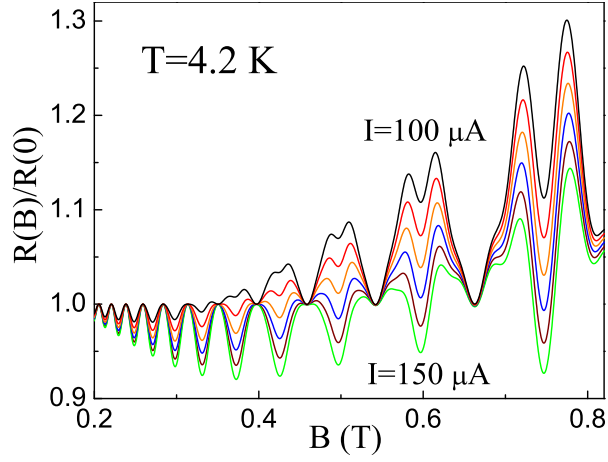}
\end{center}
\addvspace{-0.8 cm} \caption{(Color online) Evolution of the
nonlinear magnetoresistance calculated using the parameters of the
sample B when the current varies from 100 to 150 $\mu$A with the step
of 10 $\mu$A. The Gaussian model of the density of states and the
assumption of subband-independent electron-electron scattering are
used.}
\end{figure}

If the intersubband electron-electron scattering is allowed, the
magnetoresistance pictures obtained within the Gaussian model, as
well as within the SCBA model above the inversion point,
qualitatively reproduce the features we observe experimentally.
The results of calculations presented in Fig. 9 demonstrate
that varying the current in a relatively narrow range leads
to a dramatic reconstruction of the magnetoresistance oscillation
pattern.

\begin{figure}[ht]
\begin{center}\leavevmode
\includegraphics[width=8cm]{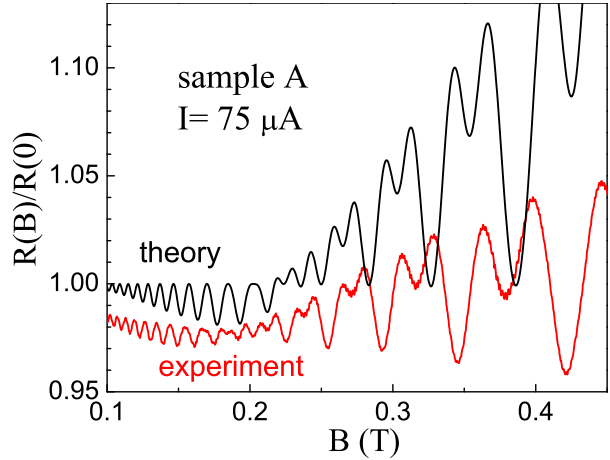}
\end{center}
\addvspace{-0.8 cm} \caption{(Color online) Comparison of the
measured and calculated nonlinear magnetoresistance in the sample A
at $T=4.2$ K and $I=75$ $\mu$A. The Gaussian model of the density of
states and the assumption of subband-independent electron-electron
scattering are used in the calculations.}
\end{figure}

Numerical calculation of magnetoresistance in the high-mobility
sample A also gives the results very similar to what we see
experimentally. To demonstrate this, we have put experimental and
calculated curves together in Fig. 10. Apart from a weak negative
magnetoresistance at low fields and a slight decrease in the MIS
oscillations frequency with increasing $B$ (the features we see in
all our samples$^{10,15}$ both in linear and nonlinear regimes), the
agreement between experiment and theory is good.

\section{Conclusions}

Investigation of nonlinear transport of 2D electrons in magnetic
fields enriches the knowledge of the quantum kinetic properties of
electron systems and of the microscopic processes responsible for
the observed modifications of the resistivity. In our work, we have
demonstrated that using double quantum well systems opens wide
possibilities for studying the nonlinear behavior. The presence of
the MIS oscillations, which modulate the quantum component of the
resistivity, allows us to investigate the current dependence of the
quantum magnetoresistance. In particular, we are able to determine
the magnetic fields $B_{inv}$ corresponding to the current-induced
inversion of the magnetoresistance. This inversion manifests itself
in a spectacular way, as a flip of the MIS oscillation pattern. We
point out that this behavior resembles recently observed$^{15}$
inversion of the MIS oscillations by the low-frequency (35 GHz)
microwave radiation. This is not surprising, because the physical
mechanism in both cases is similar. Apart from the flip of the MIS
oscillations, we have observed a wholly unexpected quantum
phenomenon, the splitting of the MIS oscillation peaks in the region
of fields above the inversion point $B_{inv}$.

We have shown that the theoretical explanation of all the observed
phenomena can be based on the kinetic equation for the isotropic
non-equilibrium part of electron distribution function. This
function oscillates with energy owing to oscillations of the density
of electron states in the magnetic field. The effect of electric
current on this function, the increase of electron diffusion in the
energy space, is equilibrated by the inelastic electron-electron
scattering. Theoretical explanation of the most of observed
phenomena is done in a simple single-harmonic approach, which
allowed us to determine the inelastic relaxation time $\tau_{in}$ by
comparison of experimental data with theory. The values of
$\tau_{in}$ for different samples are close to the theoretical
estimates of this time, and confirm the predicted$^7$ temperature
dependence $\tau_{in} \propto T^{-2}$. Thus, our data on the
inelastic relaxation time in double quantum well samples are in
agreement with the data obtained in single quantum well samples.$^5$
The description of the splitting of MIS oscillations requires a more
detailed numerical analysis including consideration of higher
harmonics of both the density of states and the distribution
function. Apart from the verification of the basic principles of the
theory of Ref. 7, this analysis demonstrates sensitivity of the
nonlinear behavior to the shape of the density of electron states
and to the details in description of inelastic scattering.
Therefore, investigation of nonlinear magnetoresistance in
relatively weak magnetic fields offers a tool for studying the
electron states and scattering mechanisms both in single and double
quantum wells.\\

This work was supported by CNPq and FAPESP (Brazilian agencies).


\begin{thebibliography}{15}

\bibitem{1}
G. Ebert, K. von Klitzing, K. Ploog, and G. Weimann, J. Phys. C {\bf
16}, 5441 (1983); M. E. Cage, R. F. Dziuba, B. F. Field, E. R.
Williams, S. M. Girvin, A. C. Gossard, D. C. Tsui, and R. J. Wagner,
Phys. Rev. Lett. {\bf 51}, 1374 (1983).

\bibitem{2}
C. L. Yang, J. Zhang, R. R. Du, J. A. Simmons, and J. L. Reno, Phys.
Rev. Lett. {\bf 89}, 076801, (2002).

\bibitem{3}
W. Zhang, H. S. Chiang, M. A. Zudov, L. N. Pfeiffer, and K. W. West,
Phys. Rev. B {\bf 75}, 041304(R) (2007).

\bibitem{4}
A. A. Bykov, J. Q. Zhang, S. Vitkalov, A. K. Kalagin, and A. K.
Bakarov, Phys. Rev. B {\bf 72}, 245307 (2005).

\bibitem{5}
J.-q. Zhang, S. Vitkalov, A. A. Bykov, A. K. Kalagin, and A. K.
Bakarov, Phys. Rev. B {\bf 75}, 081305(R) (2007).

\bibitem{6}
X. L. Lei, Appl. Phys. Lett. {\bf 90}, 132119 (2007)

\bibitem{7}
I. A. Dmitriev, M. G. Vavilov, I. L. Aleiner, A. D. Mirlin, and D.
G. Polyakov, Phys. Rev. B {\bf 71}, 115316 (2005).

\bibitem{8}
M. G. Vavilov, I. L. Aleiner, and L. I. Glazman, Phys. Rev. B {\bf
76}, 115331 (2007).

\bibitem{9}
M. G. Vavilov and I. L. Aleiner, Phys. Rev. B {\bf 69}, 035303 (2004).

\bibitem{10}
N. C. Mamani, G. M. Gusev, T. E. Lamas, A. K. Bakarov, and O. E.
Raichev, Phys. Rev. B {\bf 77}, 205327 (2008).

\bibitem{11}
See, for example, O. E. Raichev and F. T. Vasko, Phys. Rev. B {\bf
74}, 075309 (2006).

\bibitem{12}
According to our calculations, application of the current of 400
$\mu$A to the sample B at 4.2 K increases electron temperature to
6.3 K.

\bibitem{13}
Y. Ma, R. Fletcher, E. Zaremba, M. D'Iorio, C. T. Foxon, and J. J.
Harris, Phys. Rev. B {\bf 43}, 9033 (1991).

\bibitem{14}
O. E. Raichev, Phys. Rev. B {\bf 59}, 3015 (1999).

\bibitem{15}
S. Wiedmann, G. M. Gusev, O. E. Raichev, T. E. Lamas, A. K. Bakarov,
and J. C. Portal, Phys. Rev. B {\bf 78}, 121301 (2008).

\end{thebibliography}
\end{document}